\documentclass[UTF]{sigkddExp}

\begin{document}
%

\title{Modeling of the Latent Embedding of Music using Deep Neural Network}
%

\numberofauthors{1}

%


\author{
%
\alignauthor Zhou Xing, Eddy Baik, Yan Jiao, Nilesh Kulkarni, Chris Li, Gautam Muralidhar, Marzieh Parandehgheibi, Erik Reed, Abhishek Singhal, Fei Xiao and Chris Pouliot \\
       \affaddr{NIO USA, INC.}\\
       \affaddr{3200 N 1st St}\\
       \affaddr{San Jose, CA 95134}\\
       {joe.xing, eddy.baik, yan.jiao, nilesh.kulkarni, chris.li, gautam.muralidhar,marzieh.parandeh, erik.reed, abhishek.singhal, fei.xiao, chris.pouliot}@nio.com \\
}
\date{11 January 2017}

\maketitle

\begin{abstract}

While both the data volume and heterogeneity of the digital music content is huge, it has become increasingly important and convenient to build a recommendation or search system to facilitate surfacing these content to the user or consumer community. Most of the recommendation models fall into two primary species, collaborative filtering based and content based approaches.  Variants of instantiations of collaborative filtering approach suffer from the common issues of so called ``cold start" and ``long tail" problems where there is not much user interaction data to reveal user opinions or affinities on the content and also the distortion towards the popular content. Content-based approaches are sometimes limited by the richness of the available content data resulting in a heavily biased and coarse recommendation result. In recent years, the deep neural network has enjoyed a great success in large-scale image and video recognitions.  
In this paper, we propose and experiment using deep convolutional neural network to imitate how human brain processes hierarchical structures in the auditory signals, such as music, speech, etc., at various timescales. This approach can be used to discover the latent factor models of the music based upon acoustic hyper-images that are extracted from the raw audio waves of music. These latent embeddings can be used either as features to feed to subsequent models, such as collaborative filtering, or to build similarity metrics between songs, or to classify music based on the labels for training such as genre, mood, sentiment, etc. 

\end{abstract}

\section{Introduction}

While both the data volume and heterogeneity in the digital music market is huge \footnote{Spotify has hosted more than 30 million songs coupling with the 20,000 new songs being released every day, Apple music also streams 40 million songs or more presently.}, it has become increasingly important and convenient to build automated systems of recommendation and search in order to facilitate the users to locate as well as discover the relevant content. For recommendation systems, most of the models are formulated following the concept of collaborative filtering and content based methodologies. While collaborative filtering model requires substantial user feedback signals to learn or capture the user-content latent embedding, it is difficult to precisely model the latent space in a so called ``cold start" scenario when there is little signal collected from the users. Conventional content-based approaches try to solve this problem using explicit features associated with music, but limitation resides in the diversity and the level of fineness of recommendations.

\begin{figure}[!ht]
  \centering
    \includegraphics[width=0.5\textwidth]{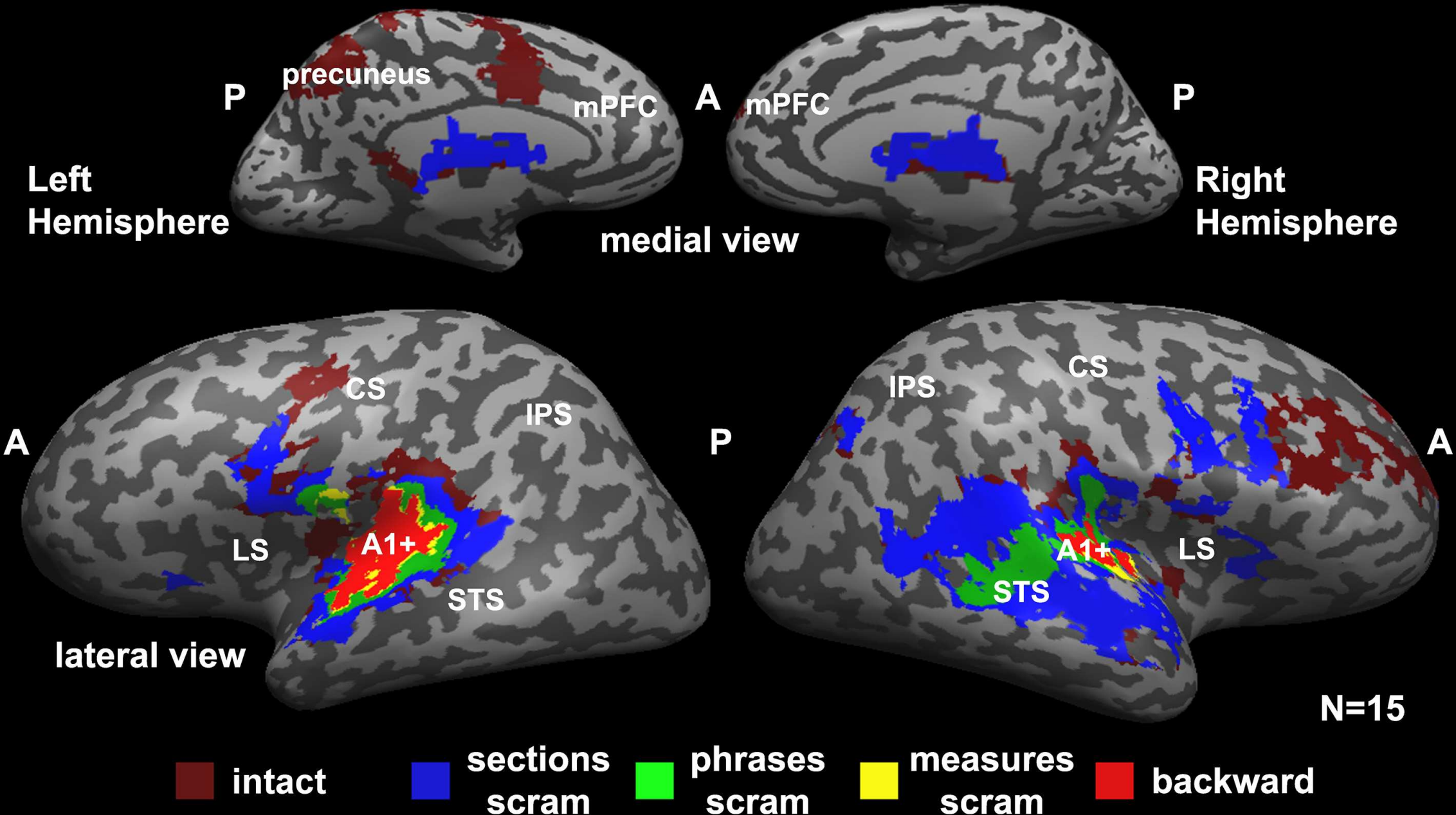}
\caption{Re-printed from [1]. Hierarchical organization of processing timescales (temporal receptive windows).}
\label{fig:neural_science}
\end{figure}

Music, as well as speech, is a complex auditory signal that contains structures at multiple timescales. Neuroscience studies \cite{Farbood2015} has shown that the human brain integrates these complex streams of audio information through various levels of voxel, auditory cortex (A1+), superior temporal gyrus (STG), and inferior frontal gyrus (IFG), etc., as shown in Fig.~\ref{fig:neural_science}. The temporal structure of the auditory signals can be well recognized by the neural network in human brain.

In recent years, deep neural network modeling has been shown to perform successful image classification and recognition tasks. 
The methodology of the localized convolution of images can also be used to recognize patterns, intensities in the music spectrogram \footnote{There are various types of spectrogram for sound or music, we combine them together to generate a so called audio ``hyper-image"}.  In this paper, we propose and experiment using deep convolutional neural network (CNN) to learn the latent models of the music based upon the acoustic data. The rationale behind this approach is that most of the explicit features associated with music, such as genre, type, rhythm, pitch, loudness, timbre, as well as latent features, for example mood, sentiment of the song, can all be embedded by various filters in the neural network. These latent embeddings of songs can be used either as features to feed to subsequent recommendation models, such as collaborative filtering, to alleviate the issues mentioned previously, or to build similarity metrics between songs, or simply to classify music into the targeted training classes such as genre, mood, etc \cite{Xing2016}. The deep learning model is experimented under CUDA computation framework using NVIDIA GTX-1080 GPU infrastructures.

\section{Description of the Dataset} 
 
 We use several Music Information Retrieval (MIR) benchmark dataset \cite{Eerola2011, Homburg2005, Mierswa2005} to train and test our deep learning model. These datasets contain raw audio files of songs that are labeled either by genre or emotion. The music genre dataset contains 1886 songs all being encoded in mp3 format. The codec information such as sampling frequency and bitrate are 44,100 Hz and 128 kb, respectively. 

\section{Extraction of the Acoustic Features} 

\begin{figure}[!ht]
  \centering
    \includegraphics[width=0.5\textwidth]{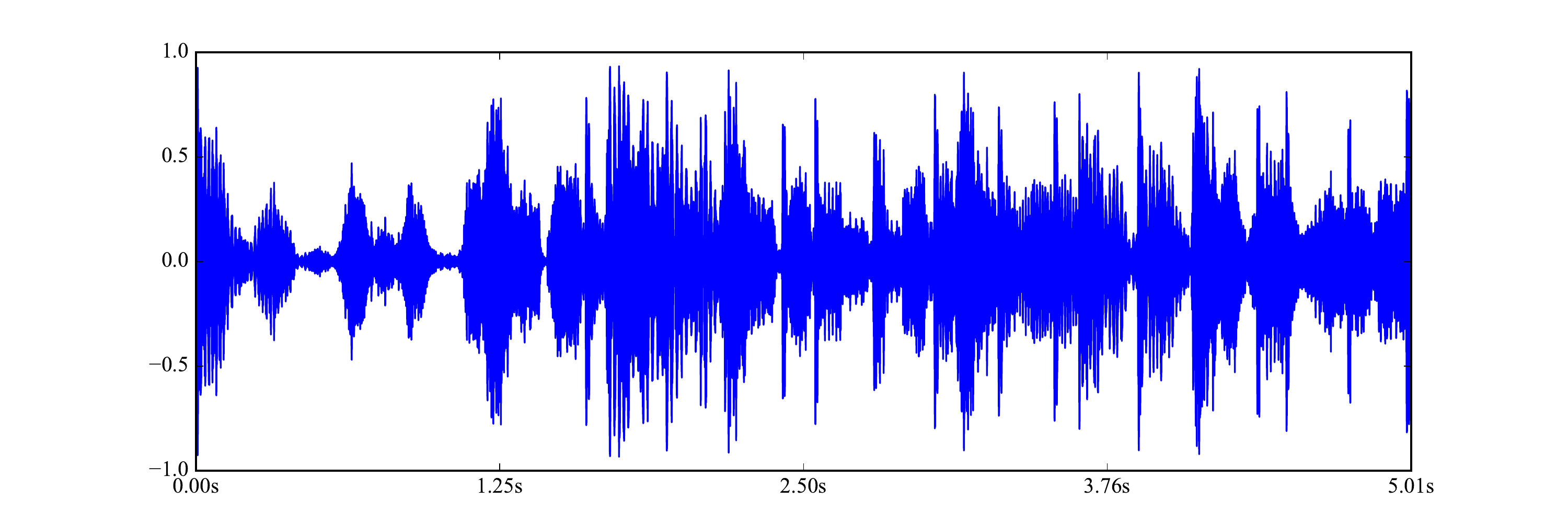}
\caption{The decoded and extracted sound wave signals from the raw mp3 or wav audio files for song ``My Humps" by ``Black Eye Peas".}
\label{fig:raw_sound_wave_signal}
\end{figure}

We use a tool, FFmpeg \cite{FFmpeg}, to decode audio files such as mp3, wav files, thus generate time series data of the sound wave. An example of the extracted sound wave signal for song ``My Humps" by ``Black Eye Peas" is shown in Fig.~\ref{fig:raw_sound_wave_signal}, only 5 seconds of snippet is shown here for illustration purpose. 

\begin{figure}[!ht]
  \centering
    \includegraphics[width=0.5\textwidth]{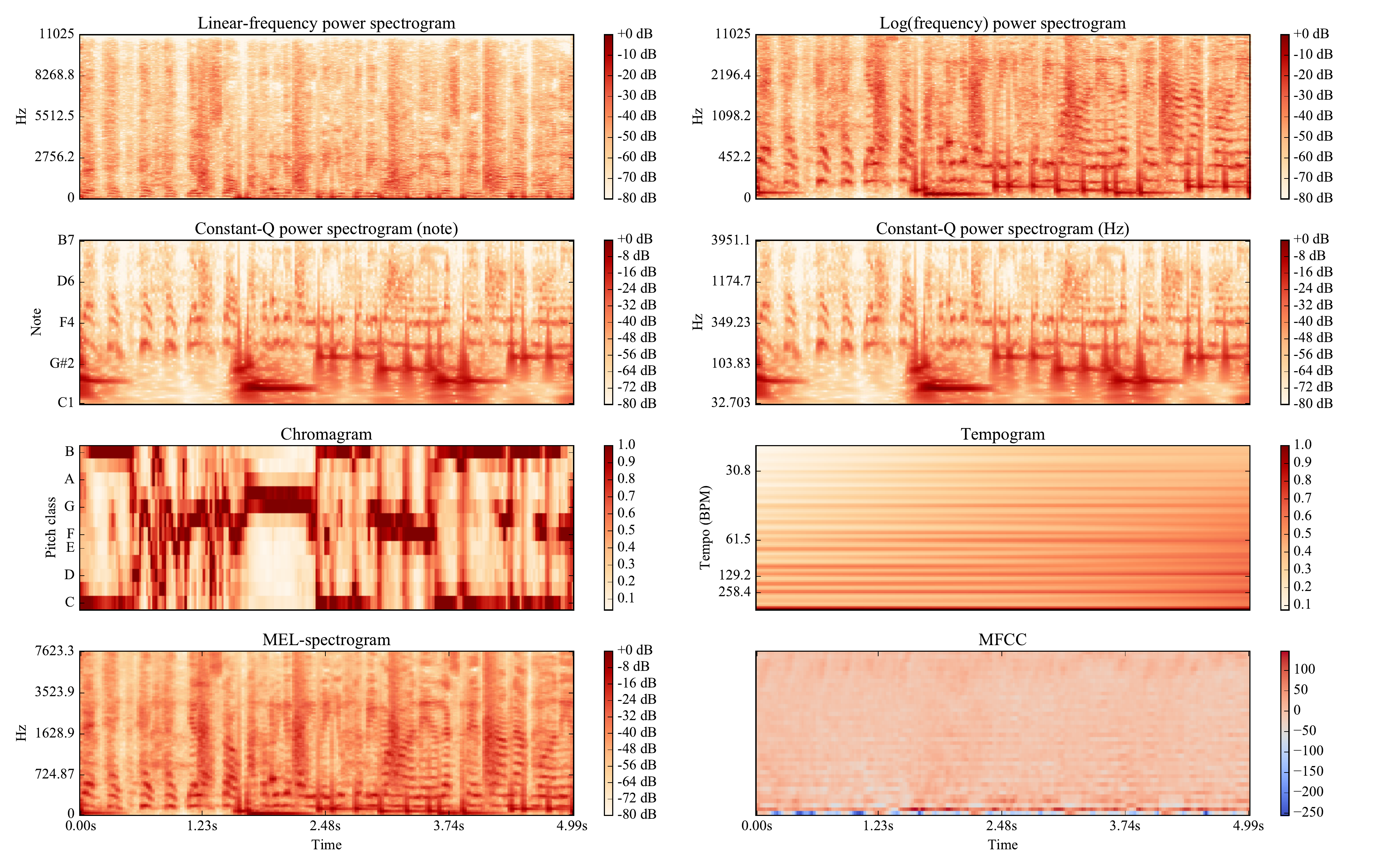}
\caption{The various time-frequency representations for song ``My Humps" by ``Black Eye Peas".}
\label{fig:hyper_image}
\end{figure}

Subsequent feature engineering involves generating time frequency representations of the sound wave. Fourier transform and constant Q-transform (CQT) can be used to generate power spectrums of the sound as shown in the top two rows of Fig.~\ref{fig:hyper_image}, in both linear and logarithmic scale, frequency and chromatic scale (music notes). Chroma-gram of the 12 pitch classes can also be constructed using short-time Fourier transforms in combination with binning strategies \cite{Bartsch2005} , as shown in the third row on the left. 
The extraction of the local tempo and beat information is implemented using a mid-level representation of cyclic tempograms, where tempi differing by a power of two are identified \cite{Grosche2010}. This cyclic tempogram is shown in the third row on the right. 

Perceptual scale of pitches such as melody spectrogram (MEL-spectrogram), mel-frequency cepstrum consisting of various Mel-frequency cepstral coefficients (MFCC) can be generated by taking discrete cosine transform of the mel logarithmic powers \cite{Logan2000} as shown in the fourth row. Especially the MFCC has been used as the dominant features in the speech recognition for some time \cite{Young1993}.

\begin{figure}[!ht]
  \centering
    \includegraphics[width=0.5\textwidth]{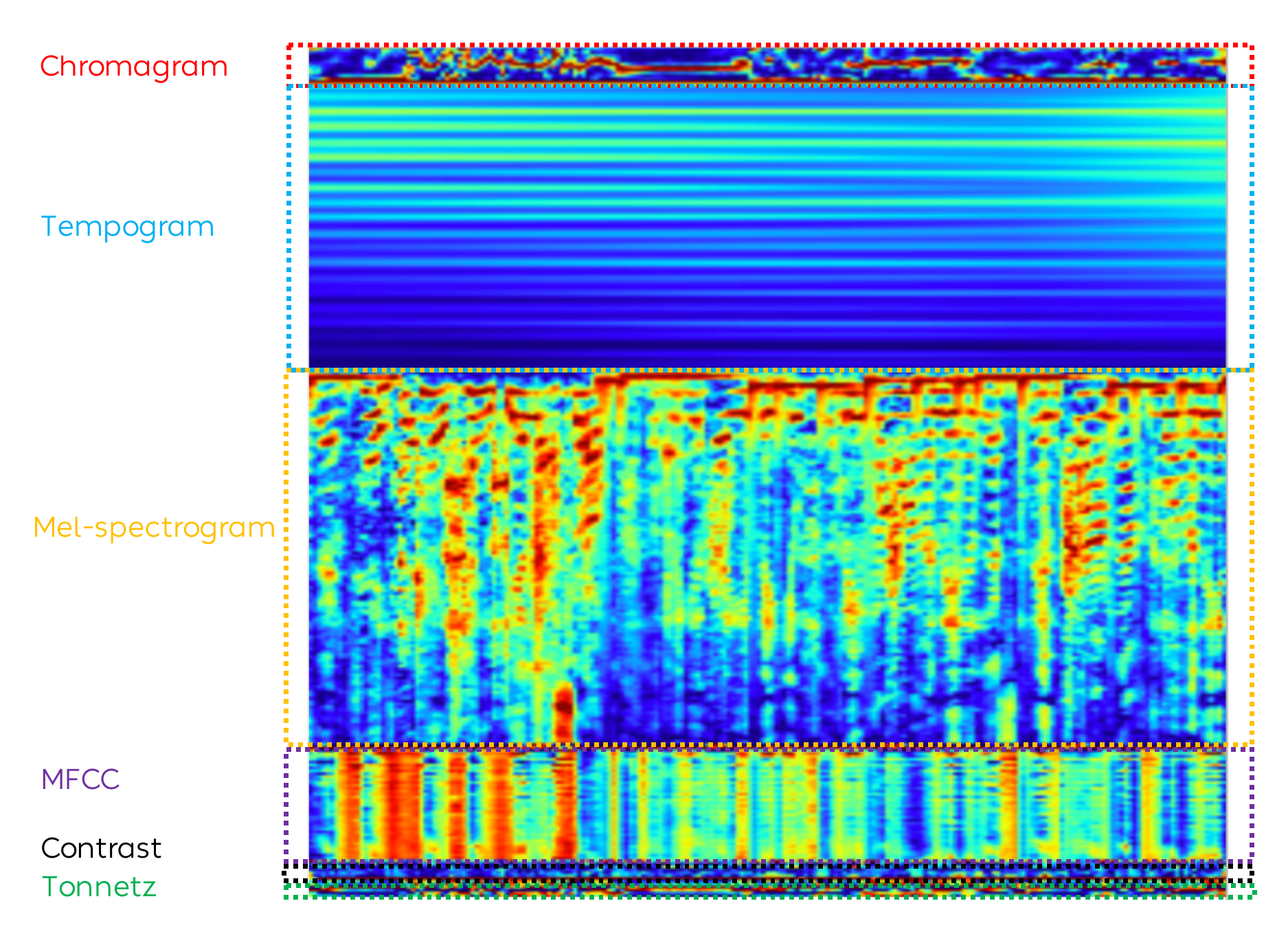}
\caption{The aggregated hyper-image that contains a lot of the acoustic information of the song such as chroma, tempo, beats, pitch, melody, cepstral information and harmonics.}
\label{fig:hyper_image_combined}
\end{figure}

We also investigate other acoustic signals such as spectral contrast \cite{Jiang2002} and music harmonics such as tonal centroid features (tonnetz) \cite{Harte2006}. In the end, we choose to combine all these the acoustic signals of chromogram, tempogram, mel-spectogram, MFCC, spectral contrast and tonnetz into a normalized ``hyper-image", as shown in Fig.~\ref{fig:hyper_image_combined}, to feed into the deep neural network for training.

\section{Architecture of the Convolutional Neural Network}

The deep neural network has been shown to perform a successful job on image classification and recognitions \cite{Krizhevsky2012}, the localized convolution to the 3-dimensional (3D) neurons has been able to capture latent features as well as other explicit features on the image such as patterns, colors, brightness, etc. Every entry in the 3D output volume, axon, can also be interpreted as an output of a neuron that picks up stimulus from only a small region in the input, and that information is shared with all neurons spatially in the same layer. All these dendrite signal, through synapse,  are integrated together to the other axon. 

\begin{figure}[!ht]
  \centering
    \includegraphics[width=0.5\textwidth]{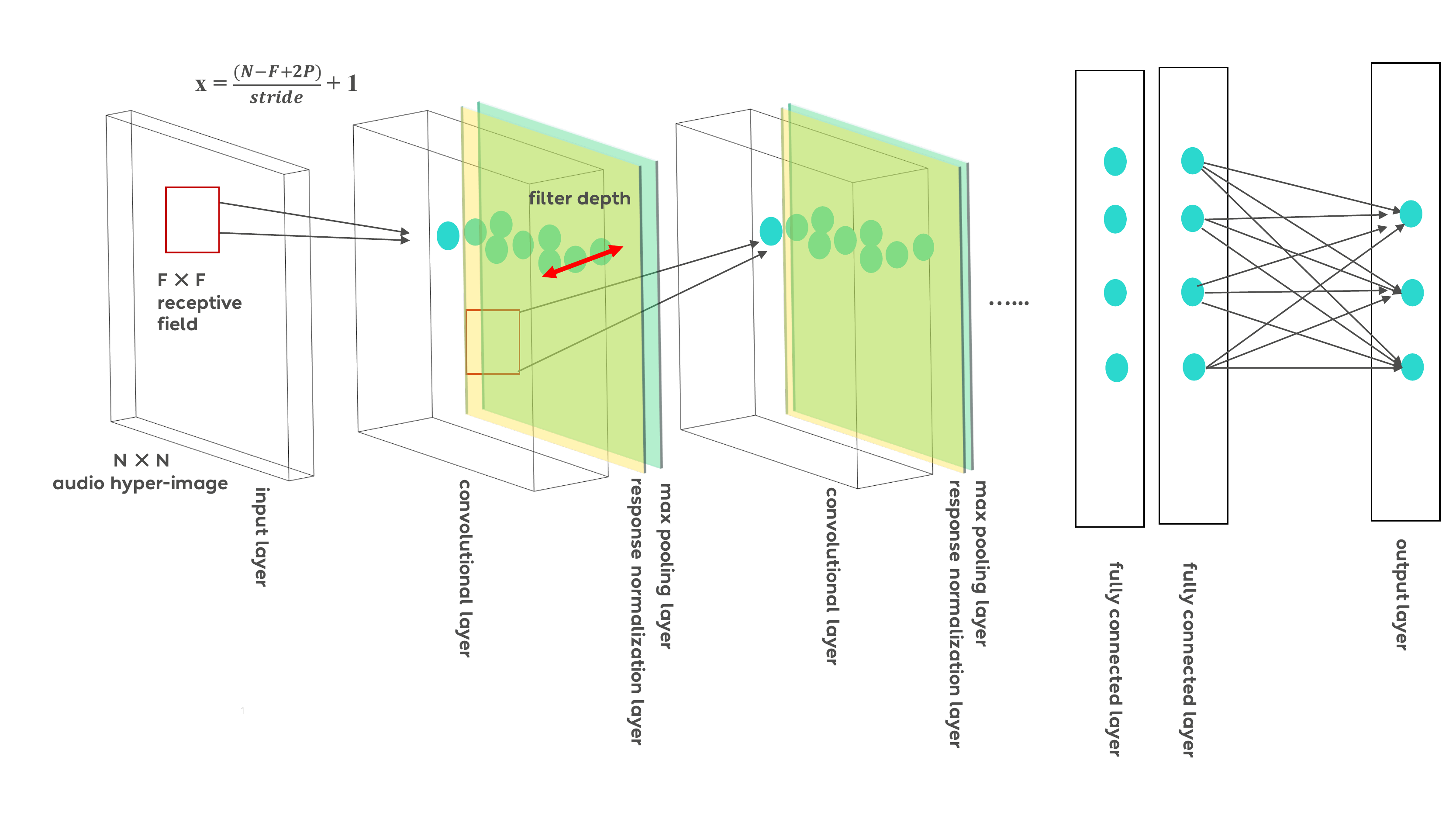}
\caption{The architecture of the convolutional neural network used to predict the music labels.}
\label{fig:cnn_architecture}
\end{figure}

We have been experimenting with various architectures of convolutional neural network in terms of the input layer dimension, convolutional layer, pooling layer, receptive filed of the neuron, filter depth, strides, as well as other implementation choices such as activation functions, local response normalization, etc. An example of the CNN architecture is illustrated in Fig.~\ref{fig:cnn_architecture}. We use stochastic gradient decent (SGD) to optimize for the weights in our neural network.

We use rectified linear units (ReLUs) to activate neurons where non-saturating nonlinearity is supposed to run faster than the saturating ones. Local response normalization (LRN) is then performed integrating over the kernel map which fulfills a form of lateral constraint for big activities between neuron outputs computed using different kernel. Max pooling is used to synchronize the outputs of neighboring groups of neurons in the same kernel map. A linear mapping is used to generate the last two fully connected layers. A softmax cross-entropy function is used to build the overall loss function.

\section{Training and Cross Validation of the Model}

A 10-fold cross validation is performed in terms of the predictions of music labels such as genres. There are 9 genres in total, alternative, blues, electronic, folk/country, funk/soul/rnb, jazz, pop, rap/hip-hop and rock, in the MIR dataset. Precisions of the predicted music label at rank 1 and 3 are shown in Table~\ref{tab:cross_validation}, where the network architecture follows the notation of I for input layer, C for convolutional layer with stride, receptive field and filter depth, L for local response normalization layer, P for max pooling layer, F for fully connected layer with number of fully connected neurons, and O for the output layer. For the way to split the training and testing sample, we have tried both selecting random snippets across all the songs for training and testing sample, as well as evenly ordered by the song title to make sure the snippets from the same song do not enter both training and testing sample simultaneously. Both methods of splitting the dataset give similar cross validation results. 

\begin{table}
\scalebox{0.7}{\parbox{.5\linewidth}{%
\begin{center}
    \begin{tabular}{ |c|c|c|c|c|}
    \hline
    fold  &  architecture         &  loss   & precision@1[$\%$] & precision@3 [$\%$] \\ 
     \hline
    0       &  IC(5,15,64)LPC(1,5,64)LPF(384)F(192)O &   2.32 & 56.9                             & 82.5 \\
    1       &  IC(5,15,64)LPC(1,5,64)LPF(384)F(192)O &   2.33 & 58.7                             & 83.1 \\
    2       &  IC(5,15,64)LPC(1,5,64)LPF(384)F(192)O &   1.26 & 55.0                             & 80.5   \\
    3       &  IC(5,15,64)LPC(1,5,64)LPF(384)F(192)O &   2.57 & 58.9                             & 84.7 \\
    4       &  IC(5,15,64)LPC(1,5,64)LPF(384)F(192)O &   2.34 &	53.7	                            & 81.5 \\
    5       &  IC(5,15,64)LPC(1,5,64)LPF(384)F(192)O &   2.32 &	57.3	                            & 85.4 \\
    6       &  IC(5,15,64)LPC(1,5,64)LPF(384)F(192)O &   2.37 &	55.2	                            & 83.2 \\
    7       &  IC(5,15,64)LPC(1,5,64)LPF(384)F(192)O &   2.36 &	57.9	                            & 82.6 \\
    8       &  IC(5,15,64)LPC(1,5,64)LPF(384)F(192)O &   1.69 & 57.3                             & 83.1  \\
    9       &  IC(5,15,64)LPC(1,5,64)LPF(384)F(192)O &   2.38 &	56.5	                            & 82.0 \\
    \hline
    \end{tabular}
\end{center}
}
}
\caption{Results of the 10-fold cross validation of the predictions of the 9 music labels. }
\label{tab:cross_validation}
\end{table}

\section{Music Embedding and Composing Music using Trained Net}

\begin{figure}[!ht]
  \centering
    \includegraphics[width=0.5\textwidth]{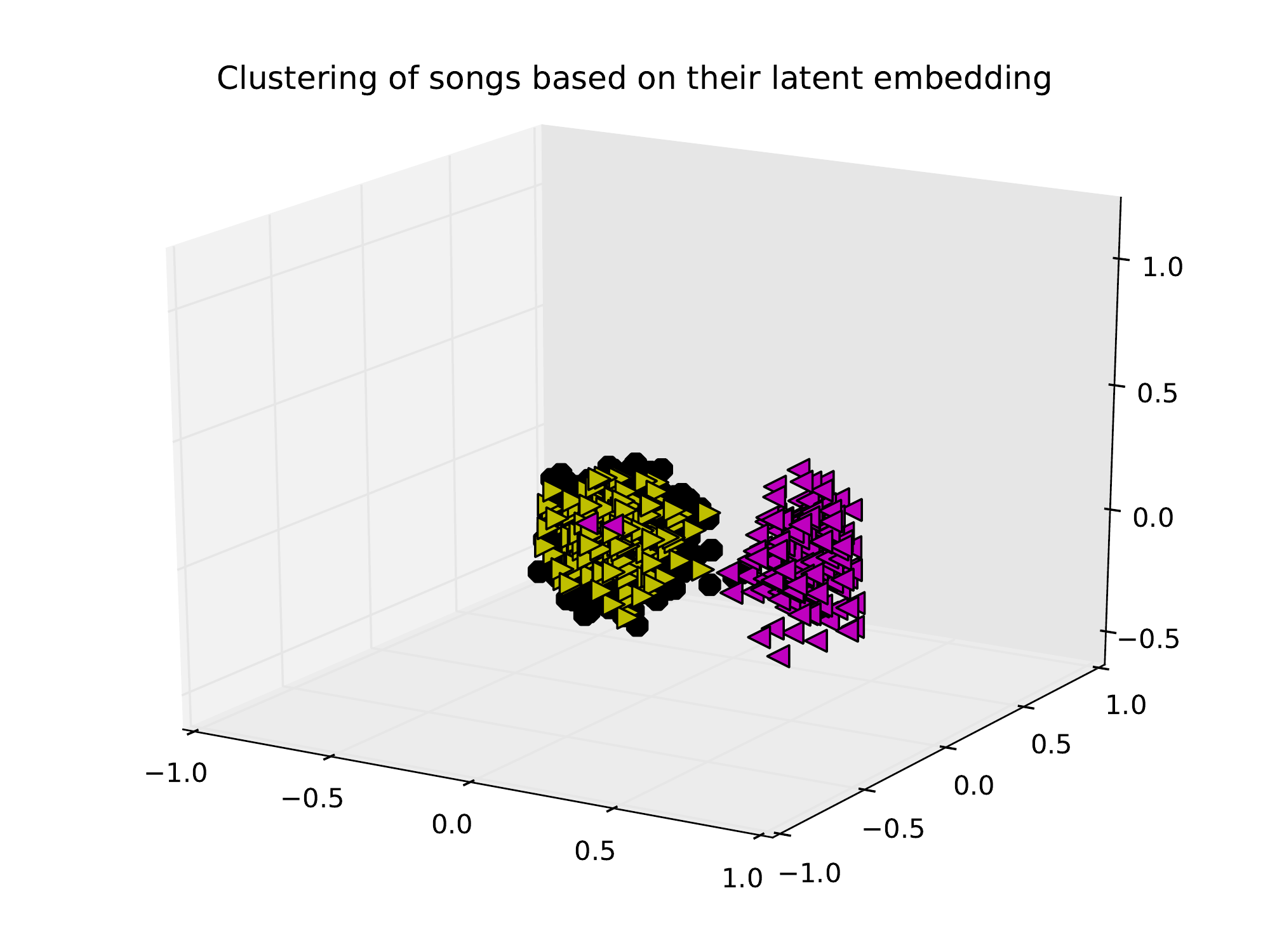}
\caption{The distribution of the song latent embedding (rank=192), in arbitrary units, after dimensionality reductions where different colors represent different song genres.}
\label{fig:song_latent_embedding}
\end{figure}

We take the values of the neurons on the last fully connected layer, just before the output layer, as the latent embedding for each song. After a dimensionality reduction using PCA we can visualize, in the 3D eigen-space, the distribution of song embedding as shown in Fig.~\ref{fig:song_latent_embedding}.

These latent vectors can be used to provide content features of each song, construct song-to-song similarity thus providing content-based recommendation. The trained weights of the neural network can also be used to compose a song, where music labels or characteristics are provided in advance, and we can perform backward propagation to artificially construct the audio hyper-image, thus the auditory signals of song. Using Artificial Intelligence (AI) techniques to compose songs, and even lyrics, would be a promising application of this particular study.

\section{Conclusions}

In this paper, we have proposed and experimented to use acoustic features of music to compose audio hyper-images, thus utilize the deep convolutional neural network to find a nonlinear mapping between these audio images and song labels. Predictions of the music labels have been tested using a 10-fold cross validation approach. We use the neurons in the last fully connected layer to embed all the songs, and these latent features can be fed to subsequent models such as collaborative filtering, factorization machine, etc., to build a recommendation system.
Further extensions of this work can be to use similar features and neural network models for voice or speech recognition, where the long-time scale of temporal structures of the auditory signals need to be captured perhaps by a recurrrent neural network structure, with the intent of the speech as the latent state variable.

%
\bibliographystyle{unsrt}

\bibliography{cnn}  

\begin{thebibliography}{10}

\bibitem{Farbood2015}
Morwaread~M. Farbood, David~J. Heeger, Gary Marcus, Uri Hasson, and Yulia
  Lerner.
\newblock The neural processing of hierarchical structure in music and speech
  at different timescales.
\newblock {\em Front. Neurosci.}, 9, 2015.

\bibitem{Xing2016}
Zhou Xing, Marzieh Parandehgheibi, Fei Xiao, Nilesh Kulkarni, and Chris
  Pouliot.
\newblock Content-based recommendation for podcast audio-items using natural
  language processing techniques.
\newblock {\em 2016 IEEE International Conference on Big Data}, 2016.

\bibitem{Eerola2011}
T.~Eerola and J.~K. Vuoskoski.
\newblock A comparison of the discrete and dimensional models of emotion in
  music.
\newblock {\em Psychology of Music}, 39(1), pages 18-49, 2011.

\bibitem{Homburg2005}
Helge Homburg, Ingo Mierswa, Bulent Moller, Katharina Morik, and Michael Wurst.
\newblock A benchmark dataset for audio classification and clustering.
\newblock {\em Proc. of the International Symposium on Music Information
  Retrieval}, pages 528--531, 2005.

\bibitem{Mierswa2005}
Ingo Mierswa and Katharina Morik.
\newblock Automatic feature extraction for classifying audio data.
\newblock {\em Machine Learning Journal}, Vol. 58, pages 127--149, 2005.

\bibitem{FFmpeg}
FFmpeg.
\newblock https://ffmpeg.org.

\bibitem{Bartsch2005}
Mark~A. Bartsch and Gregory~H. Wakefield.
\newblock Audio thumbnailing of popular music using chroma-based
  representations.
\newblock {\em IEEE Transactions on Multimedia}, 7(1), p96-104, 2005.

\bibitem{Grosche2010}
Peter Grosche, Meinard Müller, and Frank Kurth.
\newblock Cyclic tempogram - a mid-level tempo representation for music
  signals.
\newblock {\em ICASSP}, 2010.

\bibitem{Logan2000}
Beth Logan.
\newblock Mel frequency cepstral coefficients for music modeling.
\newblock {\em Proc. Int. Symp. Music Information Retrieval (ISMIR)}, 2000.

\bibitem{Young1993}
S.~J. Young, P.~C. Woodland, and W.~J. Bryne.
\newblock Htk: Hidden markov model toolkit v 1.5, cambridge university
  engineering department speech group and entropic research laboratories inc.

\bibitem{Jiang2002}
Dan~Ning Jiang, Lu~Lie, Hong~Jiang Zhang, Jian~Hua Tao, and Lian~Hong Cai.
\newblock Music type classification by spectral contrast feature.
\newblock {\em Multimedia and Expo, 2002. ICME02. Proceedings}, 2002.

\bibitem{Harte2006}
C.~Harte, M.~Sandler, and M.~Gasser.
\newblock Detecting harmonic change in musical audio.
\newblock {\em Proceedings of the 1st ACM Workshop on Audio and Music Computing
  Multimedia}, pp. 21-26, 2006.

\bibitem{Krizhevsky2012}
Alex Krizhevsky, Ilya Sutskever, and Geoffrey~E. Hinton.
\newblock Imagenet classification with deep convolutional neural networks.
\newblock {\em NIPS, 2012}, 2012.

\end{thebibliography}
%
%

\end{document}